\documentclass[aps,pra,tightenlines,floatfix,showpacs]{revtex4}
\usepackage[dvips]{graphicx}
\usepackage[english]{babel}
\usepackage{amsmath}
\usepackage{amssymb}
\usepackage{tensor}

\newcommand{\Ek}{E_{\mathbf{k}}}
\newcommand{\vk}{\mathbf{k}}
\newcommand{\vp}{\mathbf{p}}
\newcommand{\vq}{\mathbf{q}}
\newcommand{\sumk}{\sum_{\mathbf{k}}}

\newcommand{\be}{\begin{eqnarray}}
\newcommand{\ee}{\end{eqnarray}}

\newcommand{\xik}{\xi_{\mathbf{k}}}

\begin{document}

\title{Establishing the Gauge Invariant Linear Response of Fermionic Superfluids with Pair Fluctuations: A Diagrammatic approach}

\author{Yan He}
\affiliation{College of Physical Science and Technology,
Sichuan University, Chengdu, Sichuan 610064, China}
\author{Hao Guo}
\affiliation{Department of Physics, Southeast University, Nanjing 211189, China}
\date{\today}

\begin{abstract}
We present a manifestly gauge invariant linear response theory for ultra-cold Fermi gases undergoing BCS-Bose-Einstein Condensation (BEC) crossover with pair fluctuation effect included, especially in the superfluid phase, by introducing an effective external electromagnetic (EM) field. For pure BCS-type superfluids, the gauge invariance of the linear response theory can be maintained by constructing a full external EM vertex by including the fluctuation of the order parameters in the same way as the the self-energy effect is included in the quasi-particle, therefore the Ward identity (WI) is satisfied. While for the Fermionic superfluids with pairing fluctuation effect included in the quasi-particle self-energy, the construction of a gauge invariant vertex is non-trivial, since in the broken symmetry phase the effect of Nambu-Goldstone modes (collective modes) intertwines with that of the pairing fluctuation. In this paper, we find that under a suitable diagrammatic representation, the construction of such vertex is greatly simplified, which allow us to build a WI-maintaining vertex with pseudogap effects included in the superfluid phase. We focus on the $G_0G$ $t$-matrix approach to the pair fluctuations, although our formalism should also works equally well for the $G_0G_0$ $t$-matrix theory.
\end{abstract}
\maketitle

\section{introduction}
\label{intro}

Linear response theories are important theoretical tools to understand many experimental probes in condensed matter systems. Recently,
there are a lot of theoretical studies of response functions in the strongly correlated superconductors and atomic Fermi gas superfluids \cite{Stringari06,Arseev,StrinatiPRL12,HaoPRL10,HLPRA10,HLPRA12,YanPRB14}. Among these studies, a variety of approximation methods have been employed in the computation of the response functions. A natural question is how to set up a standard or criteria to improve these approximations. We believe that there does exist a natural criteria which many-body theories should satisfy as argued long time ago by Baym\cite{Baym1,Baym2}, which requires the linear response theory to respect the conservation laws of particle number, momentum and energy, etc. In our earlier works \cite{OurJLTP13,HaoJPB14}, we demonstrated that a full dressed external EM vertex can be constructed in the BCS mean field theory by treating the fluctuations of the order parameter on the same footing as the external EM disturbs. This full vertex as well as the response functions calculated from it satisfy the Ward identity or the current conservation law. Therefore the theory enjoys higher level of consistency and automatically satisfies sum rules such as longitudinal and $f$-sum rules. In this paper, we will limit our attention to the current conservation law or the gauge invariance of a linear response theory.

One important advantage of ultra-cold Fermi gases is that the inter-particle attraction can be tuned experimentally from a weak limit to a very strong limit. In this way the Fermi gas undergoes a BCS-Bose-Einstein condensation (BEC) crossover. To describe the physics of this crossover at finite temperature, many approximation methods have been put forward. One example is the $t$-matrix theory or pair fluctuation theory which successfully accounts many experimental results observed in ultra-cold Fermi gases. Although there are many theoretical works in the $t$-matrix theory, the gauge invariant linear response theory of the superfluid phase of Fermi gases with pair fluctuation effect considered is still lacking, as far as we know. The main purpose of this paper is to extend our results of the full vertex in the BCS mean field to the $t$-matrix theory.

In the following, we briefly review the $t$-matrix approach to the unitary Fermi gases. Due to the lack of small parameters, one cannot perform reliable perturbation or simple mean field calculations to describe strongly correlated systems, such as ultra-cold Fermi gases in the unitary limit or high $T_c$ superconductor. In order to capture the strong fluctuations, the $t$-matrix theory emphasizes the pairing effects by taking into account the summation of a series of ladder diagrams. The pioneer work of $t$-matrix theory appeared in Nozieres and Schmitt-Rink (NSR) \cite{NSR}, which is also known as ``$G_0G_0$" theory because the ladders is made by bare Green's functions.
However, it is widely known that $G_0G_0$ theory give rise to certain unphysical behaviors such as first order transition at $T_c$. Because of this, many improved version of $G_0G_0$ theory have been proposed\cite{Lucheroni}, which can overcome the drawbacks of NSR theory to a certain extent.

In this paper, we will focus on the $t$-matrix theory with ladder diagrams made by one bare and one full dressed Green's function, which is also called as the ``$G_0G$" theory. This approach is inspired by the early work of Kadanoff and Martin \cite{KM}. A detailed review of this theory can be found in the Ref.\cite{Ourreview}. This asymmetric choice of $G_0G$ ladder series may look strange at a first sight. But it can be shown that the this approach is more compatible with the BCS-leggett ground state\cite{KM,Maly1}. To see this, we note that the pair fluctuation of BCS theory can be treated as including virtual non-condensed pairs which are in equilibrium with the condensate of Cooper pairs. Therefore the Bose-Einstein condensation condition of the non-condensed pairs in this case can be expressed as the vanishing of ``pair chemical potential". We may interpret the $t$-matrix $t_{\textrm{pg}}(Q)$ as an amputated propagator for non-condensed pairs, then the condensation condition is equivalent to the divergence of $t_{\textrm{pg}}(0)$. From this, one can re-derive the BCS gap equation
\be
-\frac{\Delta}{g}=\sumk\Delta\frac{1-2f(\Ek)} {2\Ek}.\nonumber
\ee
Here we introduce the usual BCS quasi-particle dispersions $\Ek=\sqrt{\xik^2 + \Delta^2}$ and $\xik=\frac{k^2}{2m}-\mu$.

Below $T_c$, or in a superfluid phase of the $G_0G$ $t$-matrix theory, the self-energy can be decomposed into two parts. Aside from the usual BCS self-energy $\Sigma_{\textrm{sc}}=\Delta^2 G_0(-K)$, we also have the pseudogap self-energy which is dressed by the pair propagator or $t$-matrix as
\be
\Sigma(K)=\sum_Q t_{\textrm{pg}}(Q)G_0(Q-K)\label{sig}.
\ee
The pair propagator is given by the summation of infinite ladders made by bare and full Green's functions as
\be
t_{\textrm{pg}}(K)=\frac{g}{1+g\chi(K)},\quad\chi(K)=\sum_{Q}G_0(K-Q)G(Q).
\ee
Here $g$ is the coupling constant, $K=(i\omega_n,\vk)$ the four-momentum and the summation of four-momentum represents the summation of Matsubara frequency and momentum $\sum_K=T\sum_n\sum_{\vk}$.

There is an undetermined pseudogap self-energy appeals in the $t$-matrix which in turn determines the pseudogap self-energy. Therefore the full $G_0G$ $t$-matrix theory requires to self-consistently solve $\Sigma_{\textrm{pg}}$ from a set of coupled integral equations, which is still too complicated in practical calculations. One can employ an approximation to simplify the final result. Notice that pair condensation condition, i.e., the Thouless criterion, $t^{-1}_{\textrm{pg}} (0) = 0$, implies that
the main contribution in Eq.(\ref{sig}) comes from the vicinity of $Q = 0$. Therefore one can simplify the convolution to a multiplication
\be
\Sigma_{\textrm{pg}}(K)\approx\Big[\sum_Q t_{\textrm{pg}}(Q)\Big]G_0(-K)\equiv-\Delta_{\textrm{pg}}^2G_0(-K)
\ee
In this way, $\Sigma_{\textrm{pg}}$ takes the same form as that of the BCS self-energy, which greatly simplify the numerics. This method also provide an explicit expression for the pseudogap, therefore we will refer to it as pseudogap approximation. Although the pseudogap approximation provide convenience in numerics, it does not allow the pair propagator to carry away momentum, which makes it very difficult to satisfy the WI. In the rest of this paper, we focus on the full $G_0G$ theory without using the pseudogap approximation when constructing the full vertex. This type of approximation will only be introduced in the last step before real numerical calculations.

In constructing the WI-satisfied full vertex, we takes the similar strategy as we did in our previous works. The key point is that the fluctuation of the order parameter is dynamical and should be treated on the same footing as the external EM disturbs. When the attraction is stronger than BCS mean field theory, the calculations are far more involved than the pure BCS case because one has to take into account the effects of pair fluctuations. Thus, we first re-derive the BCS mean field results through a diagrammatic method in section \ref{BCS-diag}. It is relatively straightforward to extend this approach to the situation with pair fluctuation considered in section \ref{PG-diag}. Finally we summarize our conclusion in section \ref{conclu}.

\section{the diagrammatic proof of Ward identity in the pure BCS case}
\label{BCS-diag}

Before we present our study of the pair fluctuation case, we begin with the simpler and more familiar case about the pure BCS superfluid without pseudogaps. We will derive a gauge invariant full EM external vertex from a diagrammatic method, which is very similar to the perturbative proof of the WI in QED. The basic idea is to insert the external EM vertex into the self-energy diagram in all different possible positions, then the sum of the resulting vertex diagrams will give rise to a gauge invariant vertex.
The bare EM vertex is $\gamma^{\mu}(K+Q,K)=(1,\frac{\vp+\vq/2}{m})$ which satisfies the bare WI
\be
q_{\mu}\gamma^{\mu}(K+Q,K)=G_0^{-1}(K+Q)-G_0^{-1}(K)\label{bWI}.
\ee

\begin{figure}
\centerline{\includegraphics[clip,width=0.95\textwidth]{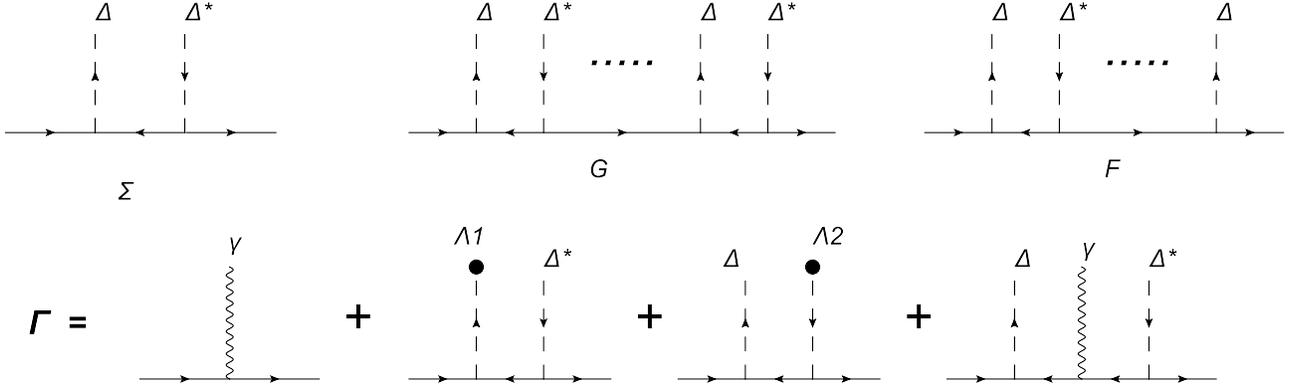}}
 \caption{The upper diagrams are BCS self-energy $\Sigma$ and full Green's functions $G$ and $F$. The thin line represents bare Green's function and the dashed line represents the paring field $\Delta$. The lower diagrams are BCS self-energy with the external EM vertex inserted in all possible ways. The wavy line represents the external EM vertex $\gamma^{\mu}$.}\label{GF}
\end{figure}

We want to insert the EM vertex at all possible places to the diagrammatic representation of the the BCS self-energy which is given by
\be
\Sigma_{\textrm{sc}}(K)=-G_0(-K)\Delta\Delta^*=\frac{|\Delta|^2}{i\omega_n+\xik}.
\ee
Here $\Delta$ and $\Delta^*$ are paring field. Comparing to the standard QED, BCS theory is more complicated because BCS vacuum contains pair condensate which breaks the $U(1)$ symmetry spontaneously. If we simply treat $\Delta$ and $\Delta^*$ as external classical fields, the $U(1)$ symmetry will be broken explicitly and there will be no current conservation. But we should remember that the paring field is actually a composite field made by two fermions and determined by the BCS self-consistent gap equation. Thus it has a more complex structure than a simple classical external field. Hence one must consider to attach the insertion of EM vertex into the paring field. Before we go to the details, it is helpful to represent the BCS self-energy by diagrams in Figure.\ref{GF}.

With the BCS-self energy, we can define the following normal and anomalous Green's functions
\be
&&G(K)=\frac{1}{i\omega_n-\xik-\Sigma_{\textrm{sc}}(K)}=\frac{i\omega_n+\xik}{(i\omega_n)^2-\xik^2-|\Delta|^2},\\
&&F(K)=\Delta G_0(-K)G(K)=\frac{-\Delta}{(i\omega_n)^2-\xik^2-|\Delta|^2}.\label{Fde}
\ee
They can also be represented by diagrams in Figure \ref{GF}. A useful property $F(K)$ is that it is an even function of both frequency and momentum $F(-K)=F(K)$.

\begin{figure}
\centering
\includegraphics[clip,width=0.95\textwidth]{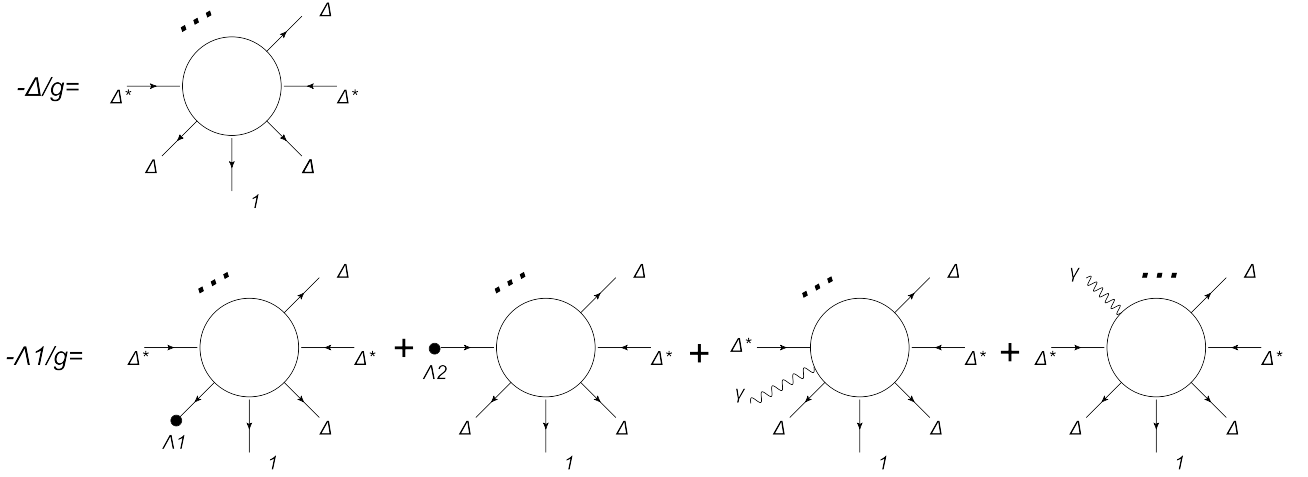}
  \caption{Diagrams for self-consistent gap equation and the gap equation with external EM vertex inserted in all possible ways. The thin line represents bare Green's function and the dashed line represents the paring field $\Delta$. The wavy line represents the external EM vertex $\gamma^{\mu}$. The black dot represents the corrected pairing field $\Lambda^{1,2}$}\label{gap}
\end{figure}

To construct a full vertex satisfying WI, we insert the EM vertex into the BCS self-energy in all possible ways. The important point is to assume that when the EM vertex has been attached to the paring fields $\Delta$ and $\Delta^*$, the resulting fields are labelled by EM-vertex-corrected pairing fields $\Lambda^{\mu}_1$ and $\Lambda^{\mu}_2$ respectively, which are unknown for now but will be determined by applying the gap equation later. Put all the above considerations together, we find the full vertex as
\be
\Gamma^{\mu}(K+Q,K)&=&\gamma^{\mu}(K+Q,K)-\Lambda^{\mu}_1\Delta^*G_0(-K)
-\Lambda^{\mu}_2\Delta G_0(-K-Q)\nonumber\\
& &-|\Delta|^2G_0(-K-Q)\gamma^{\mu}(-K-Q,-K)G_0(-K).
\ee
which can also be represented in Figure \ref{gap}.

The self-consistent gap equation defines the gap as the vacuum expectation of a pair of fermion annihilation operators
\be
\Delta=-g\langle \psi_{\uparrow}\psi_{\downarrow}\rangle=-g\sum_K F(K).
\ee
Hence the paring field acquires a phase factor under a $U(1)$ gauge transformation, which in fact cancels the other phase factor induced by the fermion operators under the same gauge transformation, therefore the BCS reduced Hamiltonian is gauge invariant. From this fact, we also see that the pairing field is dynamical, and we must consider how it is affected by the external EM field. The gap equation can be rewritten and represented by diagram as in Figure \ref{gap}.
\be
\Delta/g=\sum_n \Big[G_0(K)\Delta[-G_0(-K)]\Delta^*\Big]^nG_0(K)[-\Delta G_0(-K)].
\ee
Now we insert the EM vertex into the gap equation at all possible places as we did for the BCS self-energy. It is ether inserted to the bare Green's function or the pairing field and converts $\Delta$,$\Delta^*$ to $\Lambda^{1,2}$ as shown in Figure \ref{gap}. Therefore, we find a self-consistent equation for $\Lambda^{\mu}_{1,2}$.
\be
&&\Lambda^{\mu}_1/g=-\Lambda^{\mu}_1\sum_K G(K+Q)G(-K)+\Lambda^{\mu}_2\sum_K F(K+Q)F(K)
-2\sum_K\gamma^{\mu}(K+Q,K)G(K+Q)F(K)\label{c1},\\
&&\Lambda^{\mu}_2/g=\Lambda^{\mu}_1\sum_K F^*(K+Q)F^*(K)-\Lambda^{\mu}_2\sum_K G(-K-Q)G(K)
-2\sum_K\gamma^{\mu}(K+Q,K)F^*(K+Q)G(K).\label{c2}
\ee
Then $\Lambda^{\mu}_{1,2}$ can be solved as
\be
\Lambda^{\mu}_1=\frac{Q_{22}P_1^{\mu}-Q_{12}P_2^{\mu}}{Q_{11}Q_{22}-|Q_{12}|^2},\qquad
\Lambda^{\mu}_2=\frac{Q_{21}P_1^{\mu}-Q_{11}^*P_2^{\mu}}{Q_{11}Q_{22}-|Q_{12}|^2},
\ee
where for convenience, we have defined
\be
&&Q_{11}=\frac1g+\sum_KG(K+Q)G(-K)\label{Q11},\\
&&Q_{22}=\frac1g+\sum_KG(-K-Q)G(K)\label{Q12},\\
&&Q_{12}=Q^*_{21}=-\sum_K F(K+Q)F(K)\label{Q22},\\
&&P_1^{\mu}=-2\sum_K\gamma^{\mu}(K+Q,K)G(K+Q)F(K),\\
&&P_2^{\mu}=-2\sum_K\gamma^{\mu}(K+Q,K)F^*(K+Q)G(K).
\ee

The full vertex constructed in this way is guaranteed to be gauge invariant, yet we still need to verify this explicitly. Using the identity $G^{-1}_0(K)G(K)=1-\Delta^* F(K)=1-\Delta F^*(K)$
and Eqs.(\ref{bWI}), (\ref{Fde}),
we find the following relations
\be
& &\sum_K q_{\mu}\gamma^{\mu}(K+Q,K)G(K+Q)F(K)\nonumber\\
&=&\sum_K [1-\Delta^* F(K+Q)]F(K)-G(K+Q)\Delta G(-K)\nonumber\\
&=&-\Delta\Big(\frac1g+\sum_K G(K+Q)G(-K)\Big)-\Delta^*F(K+Q)F(K).
\ee
Similarly, we also find
\be
& &\sum_K q_{\mu}\gamma^{\mu}(K+Q,K)G(K)F^*(K+Q)\nonumber\\
&=&\sum_K G(-K-Q)\Delta^* G(K)-[1-\Delta F^*(K)]F^*(K+Q)\nonumber\\
&=&\Delta^*\Big(\frac1g+\sum_K G(-K-Q)G(K)\Big)+\Delta F^*(K+Q)F^*(K).
\ee
Comparing these two relations with Eq.(\ref{c1}) and Eq.(\ref{c2}), we find that $\Lambda_{12}^{\mu}$ have the following important properties
\be
q_{\mu}\Lambda^{\mu}_1=2\Delta,\qquad q_{\mu}\Lambda^{\mu}_2=-2\Delta^*.
\ee
Then it is easy to show that the full vertex satisfy WI.
\be
q_{\mu}\Gamma^{\mu}&=&G_0^{-1}(K+Q)-G_0^{-1}(K)+2\Sigma_{\textrm{sc}}(K)-2\Sigma_{\textrm{sc}}(K+Q)
-|\Delta|^2[G_0(-K-Q)-G_0(-K)]\nonumber\\
&=&G^{-1}(K+Q)-G^{-1}(K).
\ee
Therefore, we recover what we get in the Ref.\cite{OurJLTP13}.

\section{Gauge invariant vertex of the $G_0G$ pair fluctuation theory}
\label{PG-diag}

As we discussed in the section \ref{intro}, we must consider the effects of non-condensed pairs when treating the stronger-than-BCS attractive interaction. In this paper, we focus on the $G_0G$ pair fluctuation theory which is more compatible with the BCS-leggett ground state. However, the full $G_0G$ theory is still quite complicated due to the undetermined self-energy $\Sigma_{\textrm{pg}}(K)$ inside the full Green's function. Making use of the fact that the pair propagator is highly peaked at zero momentum, we can implement the pseudogap approximations as discussed in the section \ref{intro}, which greatly simplifies the numerical calculations. Unfortunately, the pseudogap under this approximation does not carry any momentum, which makes it impossible to satisfy the WI. Therefore, in order to make real progress, we do not introduce any approximation to the pseudogap self-energy when we construct the WI-satisfied vertex. Then the effects of the collective modes can be taking into account properly with the presence of pair fluctuation. Of course, this will make the resulting full vertex very complicated. Nevertheless, for practical calculation the pseudogap approximation will be introduced in the last step. Another complication of the $G_0G$ theory is that the full Green's function already appears in the self-energy, thus the insertion of the external EM vertex to the full Green's function requires a full vertex. Therefore, what we get finally is not a closed formula for the full vertex but a set of self-consistent equations of it.

To clarify our theory clearly, we first revisit the key ideas of the $G_0G$ pair fluctuation theory. The pseudogap self-energy is given by $\Sigma_{\textrm{pg}}(K)=\sum_Q t_{\textrm{pg}}(Q)G_0(Q-K)$ and the pair propagator is given by
\be
t_{\textrm{pg}}(K)=\frac{g}{1+g\chi(K)},\qquad\chi(K)=\sum_Q G_0(K-Q)G(Q).
\ee
We combine the BCS self-energy with the pseudogap self-energy and treat them on equal footing. The anomalous Green's function $F$ is still the expectation value of fermion pair. It takes a form that may look strange comparing to the pure BCS case, but it will be very helpful for our later discussion. The two Green's functions are expressed by
\be
&&G(K)=\frac{1}{i\omega-\xik-\Sigma_{\textrm{sc}}(K)-\Sigma_{\textrm{pg}}(K)}
=\frac{i\omega+\xik}{(i\omega)^2-\xik^2-|\Delta|^2-(i\omega+\xik)\Sigma_{\textrm{pg}}(K)},\nonumber\\
&&F(K)=\Delta G_0(-K)G(K).
\ee

Following the similar method as in the pure BCS case, we insert the EM vertex to both the BCS and the pseudogap self-energies, and find the following result find
\be
\Gamma^{\mu}(K+Q,K)&=&\gamma^{\mu}(K+Q,K)-\Lambda^{\mu}_1\Delta^*G_0(-K)
-\Lambda^{\mu}_2\Delta G_0(-K-Q)\nonumber\\
& &-|\Delta|^2G_0(-K-Q)\gamma^{\mu}(-K-Q,-K)G_0(-K)+\Lambda^{\mu}_{\textrm{pg}}(K+Q,K),
\ee
where $\Lambda^{\mu}_{\textrm{pg}}$, obtained by inserting the EM vertex to the pseudogap self-energy, is given in the Appendix, and we introduce the corrected pairing field $\Lambda_{1,2}$ again.
For convenience, we define a pseudogap vertex $\Gamma^{\mu}_{\textrm{pg}}=\gamma^{\mu}+\Lambda^{\mu}_{\textrm{pg}}$ and $\Lambda^{\mu}_{\textrm{pg}}$. This vertex itself already satisfies the WI
\be
q_{\mu}\Lambda^{\mu}_{\textrm{pg}}(K+Q,K)=\Sigma_{\textrm{pg}}(K)-\Sigma_{\textrm{pg}}(K+Q)\label{WI-pg1},\nonumber\\
q_{\mu}\Gamma^{\mu}_{\textrm{pg}}(K+Q,K)=G^{-1}_{\textrm{pg}}(K+Q)-G^{-1}_{\textrm{pg}}(K) \label{WI-pg2},
\ee
where $G^{-1}_{\textrm{pg}}(K)=G^{-1}_0(K)-\Sigma_{\textrm{pg}}(K)$ is the inverse Green's function in the normal phase.
This shows that if there is no symmetry breaking, the pseudogap theory in the normal state is indeed gauge invariant as a self-consistent many-body theory for the interacting fermion system should be.
The proof of these identities is outlined in Appendix.\ref{app}

\begin{figure}
\centering
\includegraphics[clip,width=0.95\textwidth]{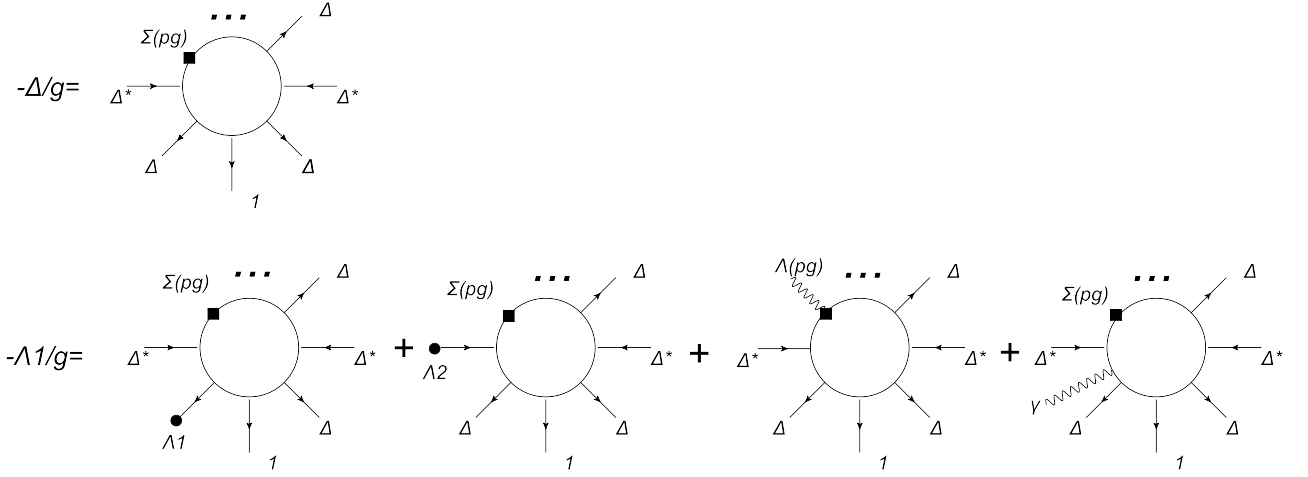}
  \caption{Diagrams for gap equation and the gap equation with external EM vertex inserted in all possible ways in the pair fluctuation case. The meaning of various symbols in the diagrams is the same as in Figure \ref{gap}. The black square represents the $\Sigma_{\textrm{pg}}$.}\label{gap1}
\end{figure}

The EM-vertex-corrected pairing field $\Lambda_{1,2}$ is again determined by inserting the EM vertex to the gap equation. With the pair fluctuation effect included, the gap equation is modified as
\be
\Delta=-g\langle \psi_{\uparrow}\psi_{\downarrow}\rangle=-g\sum_K F(K).
\ee
Since $F(K)$ contains the unknown pseudogap self-energy $\Sigma_{\textrm{pg}}$, the above gap equation becomes quite different from that of the pure BCS theory. This makes any related numerics significantly involved since the function form of $\Sigma_{\textrm{pg}}$ is unknown as discussed in the section \ref{intro}, one has to implement certain approximations in any practical calculations. However, most approximations may violate the self-consistent constraints such as the WIs. Therefore, in this paper we avoid any further approximation for now, and keep the form of gap equation unchanged to show that the WI is satisfied in our theory.

Following the same steps in the last section, We insert the EM vertex into the gap equation at all possible positions. It enters into the bare Green's function, the pairing field or the pseudogap self-energy as shown in Figure \ref{gap1}. Therefore, we find a self-consistent equation for $\Lambda^{\mu}_{1,2}$
\be
&&\Lambda^{\mu}_1/g=-\Lambda^{\mu}_1\sum_K G(K+Q)G(-K)+\Lambda^{\mu}_2\sum_K F(K+Q)F(K)\nonumber\\
&&-2\sum_K\gamma^{\mu}(K+Q,K)G(K+Q)F(K)-2\sum_K\Lambda^{\mu}_{\textrm{pg}}(K+Q,K)G(K+Q)F(K),\label{L1pg}\\
&&\Lambda^{\mu}_2/g=\Lambda^{\mu}_1\sum_K F^*(K+Q)F^*(K)-\Lambda^{\mu}_2\sum_K G(-K-Q)G(K)\nonumber\\
&&-2\sum_K\gamma^{\mu}(K+Q,K)F^*(K+Q)G(K)-2\sum_K\Lambda^{\mu}_{\textrm{pg}}(K+Q,K)F^*(K+Q)G(K).\label{L2pg}
\ee
Then $\Lambda^{\mu}_{1,2}$ can be solved as
\be
\Lambda^{\mu}_1=\frac{Q_{22}P_{1,\textrm{pg}}^{\mu}-Q_{12}P_{2,\textrm{pg}}^{\mu}}{Q_{11}Q_{22}-|Q_{12}|^2},\qquad
\Lambda^{\mu}_2=\frac{Q_{21}P_{1,\textrm{pg}}^{\mu}-Q_{11}^*P_{2,\textrm{pg}}^{\mu}}{Q_{11}Q_{22}-|Q_{12}|^2}.
\ee
Here $Q_{11}$, $Q_{12}$, $Q_{22}$ have the formally same expressions as in Eqs.(\ref{Q11}),(\ref{Q12}),(\ref{Q22}) respectively, but the real expressions are in fact more complicated since the Green's function contains the pseudogap self-energy. We have also introduced
\be
&&P_{1,\textrm{pg}}^{\mu}=-2\sum_K\gamma^{\mu}(K+Q,K)G(K+Q)F(K)-2\sum_K\Lambda^{\mu}_{\textrm{pg}}(K+Q,K)G(K+Q)F(K),\nonumber\\
&&P_{2,\textrm{pg}}^{\mu}=-2\sum_K\gamma^{\mu}(K+Q,K)F^*(K+Q)G(K)-2\sum_K\Lambda^{\mu}_{\textrm{pg}}(K+Q,K)F^*(K+Q)G(K).\nonumber
\ee

The full vertex constructed by this method must be guaranteed to satisfy WI, which can be explicitly verified. To show this, we first prove a useful identity as in the pure BCS case
\be
G^{-1}_0(K)G(K)&=&1+\frac{|\Delta|^2+(i\omega+\xik)\Sigma_{\textrm{pg}}(K)}
{(i\omega)^2-\xik^2-|\Delta|^2-(i\omega+\xik)\Sigma_{\textrm{pg}}(K)}\nonumber\\
&=&1-\Delta^* F(K)+\Sigma_{\textrm{pg}}(K)G(K).
\label{pgWI}
\ee
Then use the bare WI (\ref{bWI}) and Eq.(\ref{pgWI}), we find the following formula
\be
& &\sum_K q_{\mu}\gamma^{\mu}(K+Q,K)G(K+Q)F(K)+\sum_K q_{\mu}\Lambda^{\mu}_{\textrm{pg}}(K+Q,K)G(K+Q)F(K)\nonumber\\
&=&\sum_K [1-\Delta^* F(K+Q)]F(K)+\Sigma_{\textrm{pg}}(K+Q)G(K+Q)F(K)\nonumber\\
& &\quad-G^{-1}_0(K)G(K+Q)F(K)+[\Sigma_{\textrm{pg}}(K)-\Sigma_{\textrm{pg}}(K+Q)]G(K+Q)F(K)\nonumber\\
&=&-\Delta\Big(\frac1g+\sum_K G(K+Q)G(-K)\Big)-\Delta^*F(K+Q)F(K).
\ee
Similarly we also have
\be
& &\sum_K q_{\mu}\gamma^{\mu}(K+Q,K)G(K)F^*(K+Q)+\sum_K q_{\mu}\Lambda^{\mu}_{\textrm{pg}}(K+Q,K)G(K)F^*(K+Q)\nonumber\\
&=&\Delta^*\Big(\frac1g+\sum_K G(-K-Q)G(K)\Big)+\Delta F^*(K+Q)F^*(K).
\ee
Comparing the two above equations with Eqs.(\ref{L1pg}) and (\ref{L2pg}), we find that $\Lambda^{\mu}_{1,2}$ satisfies the following relations as in the pure BCS case.
\be
q_{\mu}\Lambda^{\mu}_1=2\Delta,\qquad q_{\mu}\Lambda^{\mu}_2=-2\Delta^*.
\ee
Now it can be straightforwardly to show that the full vertex satisfies WI as follows
\be
q_{\mu}\Gamma^{\mu}&=&G_0^{-1}(K+Q)-G_0^{-1}(K)+2\Sigma_{\textrm{sc}}(K)-2\Sigma_{\textrm{sc}}(K+Q)\nonumber\\
& &\quad-|\Delta|^2[G_0(-K-Q)-G_0(-K)]+\Sigma_{\textrm{pg}}(K)-\Sigma_{\textrm{pg}}(K+Q)\nonumber\\
&=&G^{-1}(K+Q)-G^{-1}(K).
\ee

With this full dressed vertex, we can compute the current-current correlation functions
\be
P^{\mu\nu}(Q)=2\sum_{P}q_{\mu}\Gamma^{\mu}(P+Q,P)G(P+Q)\gamma^{\nu}(P,P+Q)G(P).
\ee
This correlation function naturally satisfies the current conservation
\be
q_{\mu}P^{\mu\nu}(Q)=2\sum_{K}[G(K)-G(K+Q)]\gamma^{\nu}(K,K+Q)
=\frac{n}{m}q^{\nu}(1-\delta^{\nu0}).
\ee

As a simple application, we can show that the longitudinal sum rule and $f$-sum rule are satisfied explicitly in our theory. In component form, the current conservation can be expressed as
\be
&&\omega P^{00}(\omega,\vq)-q_j P^{j0}(\omega,\vq)=0, \label{WI2}\\
&&\omega P^{0k}(\omega,\vq)-q_jP^{jk}(\omega,\vq)=\frac{n}{m}q^k.
\label{WI}
\ee
Taking $\omega=0$ in Eq.(\ref{WI}), we find
\be
-\frac{q_j q_k P^{jk}(0,\vq)}{q^2}\equiv-P_{\textrm{L}}(0,\vq)=\frac{n}{m},
\ee
where $P_{\textrm{L}}(Q)$ is the longitudinal part of the current correlation function. Making use of the Kramers-Kronig relation, we get
\be
=\int^{+\infty}_{-\infty}\frac{d\omega}{\pi}\frac{\mbox{Im}P_{\textrm{L}}(\omega,\vq)}{\omega}
=\frac{n}{m},
\ee
which is just the longitudinal sum rule.

Similarly, Eq.(\ref{WI}) also implies that
$\omega^2 \mbox{Im}P^{00}=q_jq_k \mbox{Im}P^{jk}$,
which straightforwardly leads to
\be
\int^{+\infty}_{-\infty}\frac{d\omega}{\pi}
\big(-\omega\textrm{Im}P^{00}(\vq,\omega)\big)
=\int^{+\infty}_{-\infty}\frac{d\omega}{\pi}
\big(-\frac{\textrm{Im}P_{\textrm{L}}(\omega,\vq)}{\omega}q^2\big)
=\frac{n}{m}q^2.
\ee
This is the well-known $f$-sum rule. Therefore, by constructing the full EM vertex, we do build a fully gauge invariant linear response theory for Fermion gases with pair fluctuation effect included, where all conservation rules must be automatically satisfied. This approach is purely a theoretical attempt. In practical calculations, one need to implement suitable approximations to make the numerics simpler.

\section{conclusion}
\label{conclu}

In this paper, we have constructed the gauge invariant external EM vertex for the superfluid phase of ultra-cold Fermi gas with pair fluctuations. We achieve this result by making use of the diagrammatic method which greatly the calculation of the gauge invariant full vertex of the pure BCS theory. Based on this method, we incorporate the pseudogap vertex with the fluctuations of the order parameters. Then we can verify that the WI is satisfied by the full vertex when the contributions of the condensed pairs and the non-condensed pairs are both taken into account. With this full vertex, it is easy to construct the correlation functions, which satisfy the current conservation law and are also consistent with the sum rules. These correlation functions should be very important for the understanding of various thermodynamic or dynamical properties of Fermi gases.

Yan He is supported by NSFC under grant No. 11404228. Hao Guo is supported by NSFC under grant No. 11204032 and NSF of Jiangsu Province, China under grant No. SBK201241926.

\appendix

\section{Proof of WI for the Pseudogap vertex $\Gamma_{\textrm{pg}}$}
\label{app}

\begin{figure}
\centerline{\includegraphics[clip,width=0.9\textwidth]{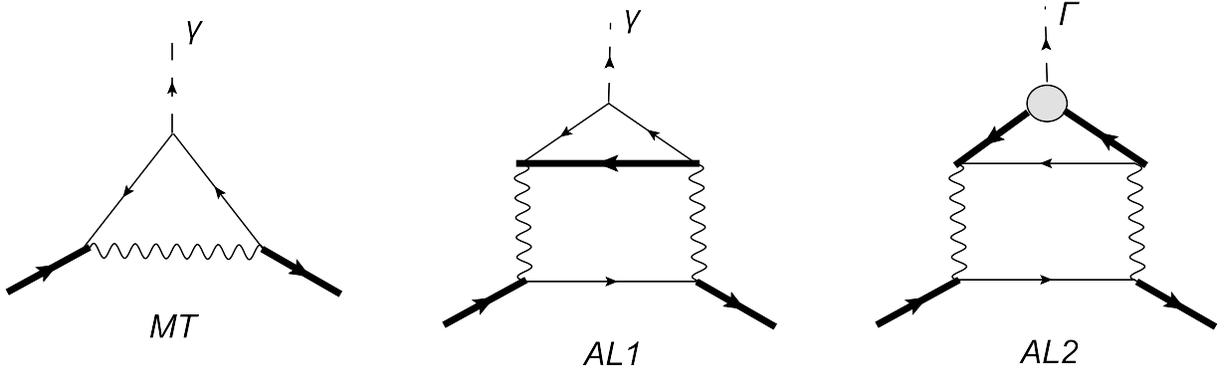}}
  \caption{Maki-Thompson diagram and two Aslamazov-Larking diagrams. The wavy line represents the pair propagator $t_{\textrm{pg}}$. The thin and thick lines represent the bare and full Green's function respectively. The dark circle represents the full vertex.}\label{MT}
\end{figure}

In this appendix, we show that the WI of the Pseudogap vertex $\Gamma_{\textrm{pg}}$ Eq. (\ref{WI-pg1})(\ref{WI-pg2}) is satisfied. By inserting the EM vertex to the pseudogap self-energy, we find 3 different diagrams. When the EM vertex is inserted to the bare Green's function, we find the familiar Maki-Thompson (MT) diagrams as shown in the Figure (\ref{MT}). When the EM vertex is inserted to the pair propagator, we find two different Aslamazov-Larking (AL) diagrams as shown in the Figure (\ref{MT}). From the diagrams, it is easy to see that the MT and AL diagrams are
\be
&&\textrm{MT}^{\mu}_{\textrm{pg}}(K+Q,K)=\sum_P t_{\textrm{pg}}(P)G_{0}(P-K)\gamma^{\mu}(P-K,P-K-Q)G_{0}(P-K-Q)\\
&&\textrm{AL}^{\mu}_{1}(K+Q,K)=-\sum_{P,L}t_{\textrm{pg}}(P)t_{\textrm{pg}}(P+Q)G_{0}(P-K)G(P-L)G_{0}(L+Q)
\gamma^{\mu}(L+Q,L)G_{0}(L)\\
&&\textrm{AL}^{\mu}_{2}(K+Q,K)=-\sum_{P,L}t_{\textrm{pg}}(P)t_{\textrm{pg}}(P+Q)G_{0}(P-K)G_{0}(P-L)G(L+Q)\Gamma^{\mu}(L+Q,L)G(L).
\ee
Then the corrected vertex is given by
\be
\Lambda^{\mu}=\textrm{MT}^{\mu}+\textrm{AL}^{\mu}_1+\textrm{AL}^{\mu}_2
\ee
Before we can show that $\Gamma_{\textrm{pg}}^{\mu}$ satisfies WI, we first show that the contribution of MT and AL diagrams will cancel each other when dotted with the external momentum. Note that we can write the pseudogap self-energy in two different ways as $\Sigma_{\textrm{pg}}(P+Q)=\sum_P t_{\textrm{pg}}(P)G_0(P-K-Q)=\sum_P t_{\textrm{pg}}(P+Q)G_0(P-K)$. Making use of this fact, we find the following identity
\be
0&=&\sum_P\big[t_{\textrm{pg}}(P+Q)G_0(P-K)-t_{\textrm{pg}}(P)G_0(P-K-Q)\big] \nonumber\\
&=&\sum_P\Big(\big[t_{\textrm{pg}}(P+Q)-t_{\textrm{pg}}(P)\big]G_0(P-K)+t_{\textrm{pg}}(P)[G_0(P-K)-G_0(P-K-Q)\big]\Big)\nonumber\\
&=&\sum_P\Big(-t_{\textrm{pg}}(P+Q)t_{\textrm{pg}}(P)\big[\chi(P+Q)-\chi(P)\big]G_0(P-K)
+t_{\textrm{pg}}(P)[G_0(P-K)-G_0(P-K-Q)\big]\Big)\label{sigma-pg}
\ee
Similarly, the pair susceptibility can also be written in two different ways as $\chi(P)=\sum_{L}G(P-L)G_0(L)=\sum_{L}G_0(P-L)G(L)$, then we find the following identity
\be
& &\chi(P+Q)-\chi(P)\nonumber\\
&=&\frac{1}{2}\sum_{L}\Big(G(P-L)\big[G_0(L+Q)-G_0(L)\big]+G_0(P-L)\big[G(L+Q)-G(L)\big]\Big)\label{chi}
\ee
Combining Eq.(\ref{sigma-pg}) and Eq.(\ref{chi}), we find
\be
0&=&-\frac{1}{2}\sum_{P,L}t_{\textrm{pg}}(P+Q)t_{\textrm{pg}}(P)G_0(P-K)G(P-L)\big[G_0(L+Q)-G_0(L)\big]\nonumber\\
& &-\frac{1}{2}\sum_{P,L}t_{\textrm{pg}}(P+Q)t_{\textrm{pg}}(P)G_0(P-K)G_0(P-L)\big[G(L+Q)-G(L)\big]\nonumber\\
& &+t_{\textrm{pg}}(P)[G_0(P-K)-G_0(P-K-Q)\big]
\ee
By applying the bare WI Eq.(\ref{bWI}) and full vertex WI, we find the following cancellation.
\be
q_{\mu}\big[\frac{1}{2}\textrm{AL}^{\mu}_{1}(K+Q,K)+\frac{1}{2}\textrm{AL}^{\mu}_{2}(K+Q,K)
+\textrm{MT}^{\mu}_{\textrm{pg}}(K+Q,K)\big]=0.
\ee
Moreover, by applying the WI Eq.(\ref{bWI}), we see that the MT vertex satisfies.
\be
q_{\mu}\textrm{MT}^{\mu}_{\textrm{pg}}(K+Q,K)&=&\sum_P t_{\textrm{pg}}(P)\big[G(P-K-Q)-G(P-K)\big]\nonumber\\
&=&\Sigma_{\textrm{pg}}(K+Q)-\Sigma_{\textrm{pg}}(K).
\ee
Collect all the above results, we find
\be
q_{\mu}\Lambda^{\mu}_{\textrm{pg}}(K+Q,K)=-q_{\mu}\textrm{MT}^{\mu}_{\textrm{pg}}(K+Q,K)
=\Sigma_{\textrm{pg}}(K)-\Sigma_{\textrm{pg}}(K+Q)
\ee
as claimed in the main text.


\end{document}